\documentclass{cimento}
\usepackage{amsmath}
\usepackage{graphicx}
\usepackage{amssymb}
\usepackage{amsfonts}
\usepackage{latexsym}

\def\beq{\begin{eqnarray}}
\def\eeq{\end{eqnarray}}
\def\ln{\,\mbox{ln}\,}
\newcommand{\Ln}{\,\mbox{Ln}\,}
\def\Det{\,\mbox{Det}\,}

\def\al{\alpha}
\def\be{\beta}

\def\ga{\gamma}
\def\de{\delta}
\def\vp{\varepsilon}
\def\ep{\epsilon}

\def\ka{\kappa}
\def\la{\lambda}
\def\na{\nabla}
\def\pa{\partial}

\def\si{\sigma}
\def\om{\omega}
\def\ph{\varphi}

\def\Ga{\Gamma}

\def\La{\Lambda}

\newcommand{\n}[1]{\label{#1}}
\newcommand{\nn}{\nonumber}
\usepackage{indentfirst}
\usepackage{mathrsfs} 

\usepackage{cite}
\usepackage{color}

\usepackage{soul}

\title{The running vacuum in effective quantum gravity
\thanks{Partially based on the talk presented by I.Sh.}}

\author{
Breno~L.~Giacchini\from{ins:x},
Tib\'erio~de~Paula~Netto\from{ins:x},
Ilya~L.~Shapiro\from{ins:y} \thanks{Also at Tomsk State
Pedagogical University.},
}

\instlist{\inst{ins:x} Department of Physics, Southern University of
Science and Technology,
\\
Shenzhen, 518055, China
\inst{ins:y}
Departamento de F\'{\i}sica,  ICE,
Universidade Federal de Juiz de Fora,
\\
Juiz de Fora,  36036-900,  MG,  Brazil}

\begin{document}

\maketitle

\begin{abstract}
We briefly review the previous works on the renormalization group
in quantum general relativity with the cosmological constant, based
on the Vilkovisky and DeWitt version of effective action. On top
of that, we discuss the prospects of the applications of this version
of renormalization group to the cosmological models with a running
Newton constant and vacuum energy density.
\end{abstract}

\section{Introduction}
\label{s1}

Nowadays, cosmology seems to be the most promising field of
application for semiclassical and quantum gravity models. One of the
most difficult problems on the interface of quantum field theory and
cosmology is the cosmological constant problem \cite{Weinberg89}.
From the field theory perspective \cite{CC-nova},  the observed
vacuum energy density may not be exactly constant because of the
remnant low-energy quantum effects. This observation led to
cosmological models with the running cosmological and Newton
constants (see \cite{PoImpo} and the last paper on the subject
\cite{Jhonny-GrCo} for a review and further references).

The main basis of the running cosmology models
\cite{CC-nova,Babic} (see also  \cite{PoImpo,DCCrun} for the
review and discussion of the theoretical side of the problem)
is the universal form of the scale-dependent vacuum energy density,
owing to the semiclassical gravitational corrections,
\beq
\rho_\La = \rho^0_\La + \frac{3\nu}{8\pi G}\,
\big(H^2 - H^2_0 \big),
\label{runCC}
\eeq
where  $\rho^0_\La$ and $H^2_0$ are the vacuum energy density and
the Hubble parameter at the reference point, e.g., in the present moment
of time. $\nu$ is a phenomenological parameter which can be
associated to the masses of the GUT particles. Regardless the
formula (\ref{runCC}) cannot be proved in the QFT framework, it
can be shown that any relevant IR running of the vacuum energy,
derived from the covariant semiclassical approach, has the form
\cite{DCCrun}. There may be other identifications, but the
systematic scale-setting procedure suggested in \cite{Babic-setting}
shows that the optimized cosmological identification of the parameter
$\mu$ is $H$. In a different physical framework
 there may be other interpretations of $\mu$.
 Using the conservation law in the models
without the particle creation from vacuum leads to the more
``traditional'' logarithmic form of the running for the Newton
constant \cite{Gruni}
\beq
G(\mu)=\frac{G_0}{1+\nu\,\log\left(H^2/H_0^2\right)}\,.
\label{runGH}
\eeq
A similar to \cite{Babic-setting} procedure of scale setting
\cite{Hrvoje} may be useful for the astrophysical applications
\cite{RotCurves} of the running Newton constant $G$. There are
several derivations of (\ref{runCC}) and (\ref{runGH}), and all
of them are (in some cases, implicitly) based on the quadratic
decoupling
of massive degrees of freedom in the IR. In other words, the
aforementioned running is due to the quantum effects of
\textit{massive} particles in the IR, according to the
corresponding theorems \cite{AC,apco}. The massless
degrees of freedom do not
produce running of the dimensional parameters, such as
$\rho_\La$ and $G$. On the other hand, it is remarkable that
(\ref{runGH}) looks exactly as the one-loop running of the
massless coupling in the Minimal Subtraction scheme of
renormalization, regardless it is derived in a very different
framework.

One can ask whether the aforementioned
universality of the running of $\rho_\La$ and $G$ holds in the full
quantum gravity (QG), when metric is a quantum field. The complete
answer to this question may be given only on the basis of a completely
consistent theory of quantum gravity, i.e., something we do not have.
However, any kind of a purely metric quantum gravity
should have massless degrees of freedom. Thus, one can assume that
in the IR the possible massive degrees of freedom (including higher
derivative ghosts) decouple and do not play a role in the running.
In this way, we arrive at the effective theory of QG, where only the
massless modes are active and, consequently, the universal effective
theory is the quantum version of general relativity (GR)
\cite{don-reviews}.

It turns out that the running of $G$ and $\La$ in the effective QG
can be achieved only by using the ``unique'' effective action of
Vilkovisky and DeWitt \cite{Vilk-Uni,DeWitt-ea}, in a way this
was considered in \cite{TV90} (previously, there was a detailed
one-loop calculation in the quantum GR with a cosmological
constant \cite{ChrisDuff}) . In the present contribution we review
the recent works \cite{UEA-contas} and especially \cite{UEA-RG},
trying to give more introductory exposition of the subject.

The manuscript is organized as follows. In the next section \ref{s2},
we present a very qualitative view on the renormalization group
running.  In section \ref{s3}, we consider how the running
in effective QG with a nonzero cosmological constant
is affected by the power counting and, in particular, explain
why the one-loop effects are strongly dominating in the effective
renormalization group equations.  Section \ref{s4} discusses the
gauge fixing ambiguity in these equations. The Vilkovisky's
construction is described in a very qualitative way in  section
\ref{s5}.
Section \ref{s7} includes a brief survey of the main
results for the running \cite{TV90,UEA-RG} in cosmology, and section \ref{s8}
aims to show that the running in effective QG is qualitatively
different from the (\ref{runCC}).
Finally, in section \ref{s9} we
draw our conclusions and discuss some of the existing
perspectives.

\section{Brief review of renormalization group}
\label{s2}

The renormalization group is a useful and economical way
to describe quantum corrections.
Thus, it is quite natural trying to use it in quantum gravity (QG).

\begin{figure}
\centerline{\includegraphics[width=88mm]{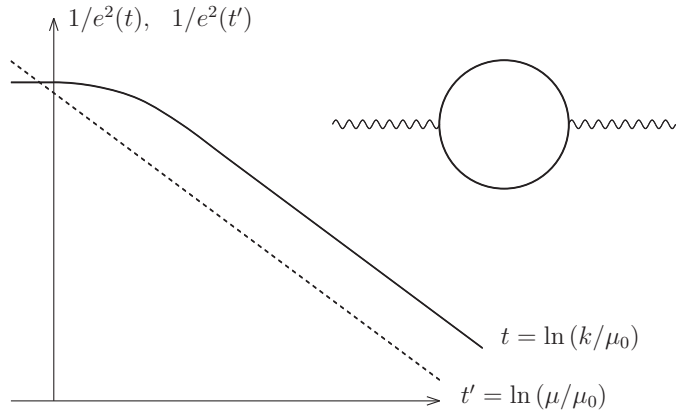}}
\caption{
The two plots illustrate the difference in the running
of the effective charge with $\log (k/\mu)$
in the MS-scheme (dashed line), and in the momentum-subtraction
scheme, where we can  see with the decoupling in the IR.}
\label{Fig1}
\end{figure}

As an example, consider a fermion loop effect in QED, where the
one-loop effective Lagrangian of a gauge vector field has the form
\beq
{\mathcal L}\,=\,-\frac{1}{4e^2}F_{\mu\nu}F^{\mu\nu}
\,+\, i\bar{\psi}\big[ \ga^\mu (\pa_\mu - A_\mu) - im \big]\psi.
\eeq
taking into account the one-loop correction, we get,
approximately,\footnote{The
precise one-loop expression can be found in many textbook as
integral representation and in \cite{QED-form} in the explicit form.}
$$
{\mathcal L}_{em}\,=\,-\frac{1}{4e^2}\,F_{\mu\nu}
\Big[1 - \be \ln\Big(-\,\frac{\Box + m^2}{\mu^2} \Big) \Big]
F^{\mu\nu}.
$$
In the IR, when (Euclidean) momentum satisfies
$k^2 \ll m^2$
this becomes an irrelevant redefinition of the charge $e$.
However, in the UV, when
$k^2 \gg m^2$
there is an effective logarithmic running
$e^2(k) \,=\,e^2_0\big(1 - \be \ln \frac{k^2}{\mu_0^2}\big)$.
An illustration of the physical running vs the one in the Minimal
Subtraction scheme with $\mu$ replacing $k$, is given in
Fig.~\ref{Fig1}.

In more details, the momentum-subtraction $\be$-function leads
to the important special cases:
\beq
&& \mbox{UV limit}
\qquad
p^2 \gg m^2 \,\,\Longrightarrow \,\,
\be_e^{1\,\,UV} \,=\,\frac{4\,e^3}{3\,(4\pi)^2}\,
+ \,{\mathcal O}\Big(\frac{m^2}{p^2}\Big)\,,\qquad\quad
\nn
\\
&& \mbox{IR limit}
\qquad
p^2 \ll m^2 \,\,\Longrightarrow \,\,
\be_e^{1\,\,IR} \,=\, \frac{e^3}{(4\pi)^2}\,\cdot\,
\,\frac{4\,p^2}{15\,m^2} \,\,
+ \,\,{\mathcal O}\Big(\frac{p^4}{m^4}\Big)\,.
\eeq
This is nothing else but the standard form of the
decoupling theorem by Appelquist and Carazzone \cite{AC}.
One can demonstrate that the decoupling works in a very
similar way in the semiclassical gravity, for both four-derivative
\cite{apco} and the Einstein-Hilbert terms
\cite{omar2013,Omar-FF}. This result gives strong, albeit
indirect support to the hypothesis of the running (\ref{runCC})
and to the respective cosmological applications.


Using renormalization group for the effective couplings
one can avoid working with explicit non-localities of the effective
action and explore interpolation between the UV and IR.
In general QG, there are both massive and
massless (graviton) degrees of freedom (see, e.g., \cite{stelle78}
and the recent textbook \cite{OUP} for a pedagogical introduction).
In the IR, all the massive modes have to decouple and we end up
with the effect of only gravitons. Owing to covariance arguments
and assuming locality of the effective QG in the IR,  the
low-energy effects of QG should be described by the quantum GR.

Second, in the pure QG based on GR, the logarithmic form factors
are much simpler because of the masslessness of the graviton. Thus,
the quantum form factors boil down to the expressions of the type
$\ln \big( -\Box/\mu^2\big)$, much simpler compared to the massive
cases. One of the consequences is that, such an effective QG is valid
in a wide interval of energy scales, i.e., between the Planck scale
in the UV and the Hubble scale in the IR, where the small
cosmological constant becomes an IR regulator for the running
owing to the massless degrees of freedom \cite{GWprT}. From the
viewpoint of the cosmological applications, this means the running
of the parameters of the action produced by quantum gravity within
the Minimal Subtraction scheme, can be applied to the whole history
of the Universe, say between the inflationary and the present-day
epochs.

Finally, in QED and in the Standard Model, the running is an
observable effect. The question is how to apply this idea to QG
and what we can learn from doing so.

\section{Gauge invariant renormalizability in QG}
\label{s3}

Let us start with a brief survey of the gauge invariant renormalizability
in quantum gravity. This is, indeed, the fundamental feature of the theory
admitting the use of effective approach \cite{Wein}.

Consider a covariant action of gravity
\beq
S = \int d^4x \sqrt{-g}\,\,{\mathcal L}(g_{\mu\nu}),
\label{gravac}
\eeq
where ${\mathcal L}(g_{\mu\nu})$ is a scalar Lagrangian that may
correspond to Einstein's GR or to
another, extended theory of gravity.
The gauge transformation
$\delta g_{\mu\nu} = R_{\mu\nu,\al}\xi^{\al}
= - \na_\mu \xi_\nu  - \na_\nu \xi_\mu$
corresponds to the infinitesimal
coordinate transformations ${x^\prime}^\mu = x^\mu + \xi^\mu$.

The gauge invariance of the classical theory means
\beq
\frac{\de S}{\de g_{\mu\nu}}\, R_{\mu\nu ,\al}\,= 0\,.
\eeq
The gauge invariant renormalizability of QG implies that
the same gauge identity is satisfied for the effective action
at the quantum level. The first proof of this fact in QG was
given by Stelle in 1977 for the fourth-derivative theory
\cite{Stelle77}. In the recent paper \cite{Lavrov-renQG} it was
generalized (using the Batalin-Vilkovisky technique) to an
arbitrary covariant theory of QG
(see also \cite{BarSib} for another consideration of the
same subject). Also, the textbook-level introduction can be
found in \cite{OUP}.

In the effective QG perspective, the gauge invariant renormalizability
means that in all loop order, all the counterterms in the quantum GR
are covariant and local expressions, that can be eliminated by the
corresponding covariant local counterterms. Thus, the form of the
counterterms is completely defined by the power counting arguments,
as described below. On top of that, the dependence on the choice of a
gauge fixing condition and the parameterization of the quantum metric
can be kept under control by the general theorems \cite{aref,volatyu},
as elaborated for QG in \cite{frts82,avram,a,JDG-QG} and, in the
general form, in \cite{Lavrov-renQG,OUP}.

In details, the gauge fixing and parameterization ambiguities in
quantum GR is as follows.
Using the Faddeev-Popov approach and the basic parameterization of
the metric,
\beq
g_{\mu\nu}(x) = \eta_{\mu\nu} + h_{\mu\nu}(x),
\label{basich}
\eeq
we arrive at the total action (using compact DeWitt's notations)
\beq
S_{tot}\,=\,S(h)
\,+\,\frac{1}{2}\,\chi^\mu Y_{\mu\nu} \chi^\nu
\,+\,{\bar C}^\al M_\al^\be C_\be,
\qquad
\mbox{where}
\qquad
M_\al^\be \,=\,\frac{\de \chi^\be}{\de h_{\mu\nu}}\,R_{\mu\nu ,\al}.
\eeq
The useful choices of gauge fixing conditions and the weight
function are the generalized harmonic (Fock - De Donder) gauge
$\chi_\mu = \pa^\nu h_{\mu\nu} - \be \pa_\mu h$
and the more general, background, gauge
$\chi_\mu = \na^\nu h_{\mu\nu} - \be \na_\mu h$,
where
$g_{\mu\nu} \to g^\prime_{\mu\nu} = g_{\mu\nu} + h_{\mu\nu}$,
while the weight operator can be chosen in the form
\beq
Y_{\mu\nu} \,=\,\frac{1}{\al}\,g_{\mu\nu}.
\eeq
In this case, there are two arbitrary gauge-fixing parameters $\al$ and
$\be$.

The most general (at the one-loop level) background parametrization
introduces more arbitrary parameters $r,\,\ga_{1,2,...,6}$
\cite{JDG-QG},
\beq
\n{bgf}
g_{\al\be} \,\,
& \longrightarrow &
\,\,  g'_{\al\be}
= e^{2 \ka r\si}
\Big[
g_{\al\be}
+ \ka \big(\ga_1\, \phi_{\al\be}
+ \ga_2\, \phi\, g_{\al\be} \big)
\nonumber
\\
&&
\quad
+ \,\,
\ka^2 \big(\ga_3\, \phi_{\al\rho}\phi^\rho_\be
+ \ga_4\, \phi_{\rho\om} \phi^{\rho\om} \,g_{\al\be}
+ \ga_5\,\phi \, \phi_{\al\be}
+ \ga_6\, \phi^2 \,g_{\al\be}
\big) \Big],
\eeq
where $g_{\al\be}$ is the background metric, while $\phi_{\al\be}$ and
$\si$ are the quantum fields, and the dimensional coupling is
$\ka^2 = 16\pi G  = 16\pi/ M_P^2$.

The power counting in QG based on GR is most easily formulated
in the case of the basic parametrization (\ref{basich}). For a diagram
with an arbitrary
number of external lines of $h_{\mu\nu}$ and the number of their
derivatives $d(G)$, the superficial degree of divergence
$\omega(G)$ is defined by the expression
\beq
\om(G)+d(G)
\,=\, \sum\limits_{l_{int}}(4 - r_I)
\,-\,4V \,+\,4\, + \,\sum\limits_{V}K_V.
\label{omega}
\eeq
The first sum is over $I$ internal lines of the diagram, $r_I$  is
the inverse power of momentum in the propagator of the given
internal line, and $V$ is the total number of vertices. The
coefficients $K_V$ represent the numbers of derivatives
in the given vertex.
In addition, there is the topological relation $L=I-V+1$, defining the number $L$ of loops.

Our interest in this contribution is the quantum GR with the
cosmological constant,
\beq
&&
S_{EH} \,=\, -\,\frac{1}{16\pi G}\int d^4x \sqrt{-g} \,(R + 2\La)\,.
\label{EH}
\eeq
In this case, $r_I=2$ for both quantum metric and ghosts and
there are $K_R=2$ and $K_\La=0$ vertices. Replacing these
numbers into (\ref{omega}), we get
\beq
\om(G) + d(G) \,=\, 2 + 2L - 2n_\La,
\label{omGR}
\eeq
where $n_\La$ is the number of zero-derivatives vertices. Formula
(\ref{omGR}) is one of the main points of our consideration. This
expression shows that

\textit{i)} Without the cosmological term and the corresponding
vertices, i.e, with $n_\La=0$, there are no logarithmic (with
$\om(G)=0$) divergences that repeat
the form of the classical Lagrangian (\ref{EH}). Instead, all the
divergences have higher and higher derivatives, when the number
of loops $L$ is growing.

\textit{ii)} When $\La \neq 0$, the vertices with growing number
$n_\La$ compensate the loop number $L$. As a result, in all loop
orders we meet both cosmological constant and Einstein-Hilbert
type counterterms.

\textit{iii)} The higher loop contributions to the logarithmic
divergences are always multiplied by the factors of $(\ka^2\La)^L$,
or, equivalently, of the factors of $(\La/M_P^2)^L$.  It is fairly
easy to write down renormalization relations for $\ka^2$ and
$\La$ and use them to derive the beta functions for these
parameters. Thus, we can conclude that these beta functions
are given by the powers series in $\La/M_P^2$.

In all known theories, at all energy scales, we have
$|\La/M_P^2| \ll 1$. Then, the beta functions for both Newton
constant $G$ and the vacuum energy $\rho_\La = \La/(8\pi G)$
have the dominating one-loop parts, while the higher loops are
producing very small contributions to the running of the respective
parameters. Thus, the main question is whether one can define
the one loop beta functions in the non-renormalizable theories
in a sufficiently sound way.

\section{Gauge-fixing dependence}
\label{s4}

The main obstacle to formulate the renormalization group in the
effective framework described above is that, for the QG based
on GR with the cosmological constant, we meet a serious problem
with the gauge- and parametrization ambiguity of the counterterms.
The gauge-fixing dependence at the one-loop level was
explored directly in several works starting from the pioneer work
\cite{KTT} (see also more detailed calculations in
\cite{Kalmykov:1995fd,DalMaz,Kalmykov:1998cv}). The dependence
on the parametrization was explored in \cite{FirPei,JDG-QG} and,
more recently, in \cite{effQG} devoted to the effects of QG on
the motion of test particles, as suggested in \cite{DalMaz}. Let us
note that this motion does not depend on both gauge-fixing and
parametrization for the reasons explained below.

Indeed, one can explore the mentioned ambiguities without explicit
calculations. The point is that, in quantum GR, as in any other gauge
theory, the on-shell quantities are well-defined, invariant and
universal \cite{aref,volatyu,Lavrov-renQG}. At the one-loop level,
the on-shell condition has to be derived from the classical equations
of motion, in our case for the theory (\ref{EH}). Let us see how it
works in practice.

According to the power counting (\ref{omGR}),
the one-loop divergences have the form
\beq
\Ga^{(1)}_{div} =\frac{1}{\ep}\,
\int d^4x\sqrt{-g}\{
c_1\,R_{\mu\nu\al\be}^2 + c_2R_{\al\be}^2 + c_3R^2
+ c_4  {\Box}R + c_5R + c_6  \},
\label{1loop}
\eeq
where $c_k$ may depend on the gauge-fixing and/or parametrization
parameters $\al_i$. Taken the Weinberg's theorem
\cite{Weinberg1960}, the divergences correspond to the local
functional, thus  $c_k$ are $c$-numbers and $\Ga^{(1)}_{div}$
is local. Thus, for the two different sets $\al_i$ and $\al^{0}_i$
\cite{JDG-QG},
\beq
&&
\Ga^{(1)}_{div}(\al_i) \,-\, \Ga^{(1)}_{div}(\al^{0}_i)
\,\,=\, \frac{1}{\ep}\,
\int d^4x\sqrt{-g}\,\Big(
b_1 R_{\mu\nu} + b_2 Rg_{\mu\nu}
+ b_3 g_{\mu\nu} \La
\Big)\,\vp^{\mu\nu}\,,
\label{aps}
\eeq
where $b_k=b_k(\al_i)$ depend on the gauge-fixing and
parametrization of quantum metric. The classical on-shell
is defined by the expression
\beq
\vp^{\mu\nu}\,=\,\frac{1}{\sqrt{-g}}\,\frac{\de S}{\de g_{\mu\nu}}
\,\,\,\sim \,\,\,R^{\mu\nu}-\frac12\,\big(R+2\La\big)g^{\mu\nu}.
\label{eps}
\eeq
Replacing (\ref{eps}) into (\ref{aps}) and comparing the result
with (\ref{1loop}), the unique two invariant combinations are
\beq
c_1
\qquad
\mbox{{ and}}
\qquad
c_{\rm inv}\,=\,c_6 - 4\La c_5 + 4\La^2 c_2 + 16\La^2 c_3\,.
\label{invs}
\eeq

This situation gives rise to the well-defined on-shell
renormalization group equation for the dimensionless combination
$\la = 16\pi G\La$ of the parameters $G$ and $\La$ \cite{frts82},
\beq
\mu \frac{d\la}{\mu}\,=\, - \frac{29}{5}\,\frac{\la^2}{(4\pi)^2}.
\eeq
However, (\ref{invs}) means that there are no well-defined
individual running of the physically interesting terms, e.g.,
$\rho_\La$, \  $R$, and  $R^2$ in the effective framework based
on the conventional perturbative QG. In order to apply the scheme
described in the previous section, we need a qualitatively new input.
Resolving the problem of gauge-fixing and parametrization ambiguity
enables one to use renormalization group in the low-energy QG, that
should be an important addition to the existing results in effective
approach \cite{don-reviews,Burgess}.

\section{Vilkovisky--DeWitt (VdW) effective action in QG}
\label{s5}

A possible solution of the problem of ambiguity can be based on the
Vilkovisky--DeWitt (VdW) scheme for constructing the ``unique''
effective action. The scheme was suggested by Vilkovisky in
\cite{Vilk-Uni} and extended by De Witt \cite{DeWitt-ea} (see also
further discussions in
\cite{Fradkin:1983nw,Kunstatter:1986qa,Huggins:1987zw}). The
formalism was applied to quantum gravity in \cite{bavi85} and, in
case of the QG theory with a cosmological constant, \cite{TV90}, and
more recently in \cite{UEA-contas,UEA-RG}. In particular, \cite{TV90}
and \cite{UEA-RG} were addressing the renormalization group issue,
which we are reviewing here.

The original proposal of \cite{Vilk-Uni} makes the one-loop
divergences independent on the gauge-fixing and parametrization
of the quantum metric. In brief, the construction looks as follows.
Consider the parametrization dependence in a non-gauge quantum
field model. The one-loop expression for the theory with the
classical action $S[\Phi_i]$ has the form
\beq
{\bar \Ga}^{(1)}\,=\,\frac{i}{2}\,\Ln\Det\,S^{\prime\prime}_{ij},
\qquad
\mbox{where}
\qquad
S^{\prime\prime}_{ij} = \frac{\de^2 S}{\de \ph_i\,\de \ph_j}\,.
\label{1loopEA}
\eeq
Changing the variables $\ph_i=\ph_i(\ph^\prime_k)$, we meet
\beq
{\bar \Ga}^{(1)} = \Ln\Det
\Big(\frac{\de^ 2 S}{\de \ph^\prime_l\,\de \ph^\prime_k}\Big)
 = \Ln\Det \Big(S^{\prime\prime}_{ij}\cdot
\frac{\de \ph_i}{\de \ph^\prime_k}
\,\frac{\de \ph_j}{\de \ph^\prime_l}+
\frac{\de S}{\de \ph_i}
\frac{\de^2 \ph_i}{\de \ph^\prime_l\,\de \ph^\prime_k}\Big).
\label{1looptrans}
\eeq
One can see, that the two one-loop results coincide
on the classical equations of motion (on shell), when
\beq
\vp^i\,=\,\frac{\de S}{\de \ph_i}=0,
\eeq
but are different off-shell. The main idea of the ``unique'' effective
action is to change the definition (\ref{1loopEA}), replacing the
usual variational derivatives by the covariant  variational
derivatives, constructed on the basis of a metric in the space
of quantum fields. Then the transformation in (\ref{1looptrans})
becomes of the tensorial type and there is no off-shell difference.
It turns out that this scheme really works and the situation is
qualitatively the same in the gauge theories, regardless there are
serious technical and conceptual complications. The gauge-fixing
dependence can be regarded as a particular type of reparametrization
of quantum fields. One of the main questions is whether the choice
of the metric in the space of quantum fields can affect the result
and whether this leads to a new ambiguity (see, e.g.,
\cite{Lavrov:1988is}).


\section{Running of $G$ and $\La$ based on the
Vilkovisky effective action}
\label{s7}

For the effective QG based on Einstein's GR with the
cosmological constant, the prescription described above gives
the one-loop divergences \cite{TV90,UEA-contas}
\beq
\bar{\Ga}^{(1)}_{\text{div}}
\,=\,  - \frac{1}{\ep}
\int  \text{d}^4 x \sqrt{-g} \,
\left\{ \frac{121}{60} C^2
- \frac{151}{180} E + \frac{31}{36} R^2
+ 8 \La R + 12 \La^2\right\}.
\label{1loopdivs}
\eeq
This, completely invariant, result enables us to construct the
renormalization group equations  \cite{TV90,UEA-RG}
\beq
&&
\mu\frac{\text{d}}{\text{d}\mu}\bigg(\frac{1}{16\pi G}\bigg)
\,=\,
\frac{8\La}{(4\pi)^2}\,,
\nn
\\
&&
\mu\frac{\text{d} \La}{\text{d}\mu}
\,=\,
- \frac{2}{(4\pi)^2}\,16\pi G \La^2\,.
\label{RGs}
\eeq
The solutions can be easily found in the form
\beq
&&
G(\mu)
\,=\,
\frac{G_0\,\,\,}{
\big[1+ \frac{10}{(4\pi)^2}\,\ga_0\ln \frac{\mu}{\mu_0}\big]^{4/5}}\,,
\nn
\\
&&
\La(\mu)
\,=\,
\frac{\La_0\,\,\,}{
\big[1+ \frac{10}{(4\pi)^2}\,\ga_0\ln \frac{\mu}{\mu_0}\big]^{1/5}},
\label{runGLa}
\eeq
where $\ga_0=16\pi G_0 \La_0^2$ is a dimensionless combination of
the initial values of the two running charges. In this way, we arrive
at the well-defined running of the Newton and cosmological
constants between the Planck and Hubble scales in effective
low-energy QG. According to what we explained in the
previous sections, the main features of this running are as follows.

\textit{i)} \ Universality, that is the independence of the gauge
fixing and, in general, of the parametrization of quantum fields.

\textit{ii)} \
Let us remember that the higher-loop corrections to the equations
(\ref{RGs}) are suppressed by the powers of $\frac{\La}{M_P^2}$.
In the present-day Universe this quantity is of the order of
$10^{-120}$, but even in the inflationary epoch it is at least
$10^{-12}$. Therefore, these equations describe an effectively
exact running, independent on the loop expansion.

Taken that it is certain that below the Planck scale
all possible massive gravitational degrees of freedom decouple, the
unique assumption that leads to (\ref{RGs}) and (\ref{runGLa}) is
that the Vilkovisky's effective action is a ``correct'' prescription in
QG. In our opinion, this is a fundamental assumption that cannot
be verified in a purely theoretical framework.

Looking beyond the lowest-order effective framework, the full action
of gravity includes higher derivative terms, and can be cast into the
form \cite{highderi,SRQG-betas}
\beq
&&
S_{\text{tot}}
\,=\,  \int \text{d}^4 x \sqrt{-g} \,
\Big\{
-  \frac{1}{\ka^2}\big(R+2\La\big)
-  \frac{1}{2\la}C^2
+ \frac{1}{2\rho}E
-  \frac{1}{2\xi}R^2
\nonumber
\\
&&
+ \sum_{n=1}^{N}\Big[
\om_{n,C} C \square^n C
+ \om_{n,R} R \, \square^{n} R\Big]
+ {\mathcal O}(R_{\dots}^3)
\Big\},
\nonumber
\eeq
where $\la$, $\rho$, $\xi$ are the dimensionless parameters of the
action and ${\mathcal O}(R_{\dots}^3)$ stand for the third and
higher order in curvatures terms.
The renormalization of the higher derivative terms in the
effective QG performs similar to the renormalization of the vacuum
action of gravity in the semiclassical approach.

Let us consider only one example and refer the reader to
\cite{UEA-RG} for a more complete exposition, including
an example of the effectively exact running of the six-derivative
coefficient, derived from the two-loop calculations \cite{GorSag}.

At the one-loop
level we meet an  exact beta function for the fourth-derivative term
\beq
\mu\frac{d\la}{d\mu}\,=\,-\frac{a^2}{(4\pi)^2 } \, \la^2\,,
\qquad
a^2
\,=\,
a_{\mbox{\tiny QG}}^2\,+\, \frac{1}{5} \,+\, \frac{N_f}{10},
\qquad
a_{\mbox{\tiny QG}}^2=\frac{121}{30}.
\label{run4der}
\eeq
where we included the one-loop semiclassical contributions for
the sake of completeness and $N_f$ is the number of fermion
fields, active (not decoupled) at the corresponding energy scale.
The coefficient $a_{\mbox{\tiny QG}}^2$ is taken from the
expression (\ref{1loopdivs}) and is gauge- and
parametrization-independent.

\section{On the running vacuum cosmology in the effective QG}
\label{s8}

We intend to discuss the cosmological applications of the
running in effective QG elsewhere and now give just a few
general qualitative observations.

Assuming small cosmological constant in the present-day Universe,
we obtain
\beq
\frac{10}{(4\pi)^2}\,|\ga_0|
\,\, \ll \,\, 1.
\label{ineq}
\eeq
On the other hand, we can assume the standard identification of the
renormalization group parameter $\mu$ with the Hubble parameter,
$\mu = H$, in cosmology \cite{CC-nova,Babic-setting,DCCrun}.

In this way, the running solutions (\ref{runGLa}) lead to the
following relations
\beq
\frac{1}{G(H)}
\,=\,
\frac{1}{G_0}
\bigg[1\,+\, \frac{8}{(4\pi)^2}\,\ga_0\,\ln \Big(\frac{H}{H_0}\Big)\bigg]
\label{sol-ka-1}
\eeq
and
\beq
\rho_\La(H)
\,=\,
\rho^0_{\La}\,
\bigg[1 + \frac{6}{(4\pi)^2}\,\ga_0\,\ln \Big(\frac{H}{H_0}\Big)\bigg].
\label{sol-La-1}
\eeq
At this point we note that the running in the effective QG framework
is different from the one in the effective semiclassical approach
(\ref{runGH}) and (\ref{runCC}).

As we have already mentioned earlier, the cosmological and other
implications of the running relations (\ref{sol-ka-1}) and
(\ref{sol-La-1}) will be discussed elsewhere. Let us just make
one important observation. In the semiclassical model of running
without the energy exchange between vacuum and matter sectors,
the (\ref{runCC}) can be linked to the running of $G$ by the
conservation law \cite{Gruni}, that is also an important part of
the scale-setting procedure \cite{Babic-setting} and \cite{Hrvoje}.
It is clear that this is \textit{not} the case for eqs. (\ref{sol-ka-1})
and (\ref{sol-La-1}). The reason is that, along with these two
relations, there are also the dependence on $\mu$ and,
consequently, on $H$, in the infinite set of higher  derivative
terms, similar to (\ref{run4der}). Thus, the conservation
equation should include all these terms too.

On the other hand, the higher derivative terms are strongly
Planck-suppressed in the IR (compared to $M_P$), that
is during all the history of the Universe. As a result, we
meet a challenging task to construct the cosmological model
with a running vacuum which may violate the conservation
equation.

\section{Conclusions and discussions}
\label{s9}

The construction of QG theory which is not restricted to the
IR region, is not possible without higher derivative terms.
However, at the energies much lower than the Planck scale,
the massive modes of gravity decouple and the QG may be
described by the quantum GR. We have seen that this fact,
together with the Vilkovisky and De Witt version of the
``unique effective action'' are sufficient to describe the  running
of cosmological, Newton constant and even of all the parameters
of the higher derivative terms in the UV complete action of
gravity.

Assuming the Vilkovisky-DeWitt version of unique effective
action, we arrive at the well-defined renormalization group
equations, which turn out exact, in the sense they are free from
the higher-loop corrections. The ambiguity in the definition of
the metric in the space of the fields can be fixed by using a
reasonable set of assumptions.

Concerning the cosmological and astrophysical applications of
the solutions (\ref{sol-ka-1}) and (\ref{sol-La-1}), the perspectives
are quite extensive. The main point is that we have, for the first
time, the well-defined running of $\rho_\La$ and $G$ in the effective
QG setting, that is based \textit{only} on the three assumptions:
\textit{\it 1)} Accepting the Vilkovisky-DeWitt effective action;
\textit{\it 2)} Assuming that the massive modes in QG decouple
in the IR, that is a standard feature to believe it; \textit{\it 3)}
Identification of scale $\mu \sim H$ in cosmology and some
working identification (see, e.g., \cite{RotCurves,Hrvoje}) in
astrophysics. In our opinion, those are, by far, the weakest and
most ``natural'' assumptions one can have in the running vacuum
models.

\section*{Acknowledgments}
The work of I.Sh. is partially supported by Conselho Nacional de
Desenvolvimento Cient\'{i}fico e Tecnol\'{o}gico - CNPq under the
grant 303635/2018-5, by Funda\c{c}\~{a}o de Amparo \`a Pesquisa
de Minas Gerais - FAPEMIG under the grant PPM-00604-18, and by
Ministry of Education of Russian Federation under the project
No. FEWF-2020-0003.



\end{document}